\documentstyle[12pt]{article}
\begin{document}
\rightline{Preprint DFPD 94/TH/66, December 1994}

\vspace{0.6cm}

\begin{center}
{\bf DYNAMICS OF GENERALIZED COHERENT STATES}
\end{center}

\vspace{0.6cm}

\begin{center}
{\large Salvatore De Martino \footnote{Electronic Mail:
demartino@vaxsa.dia.unisa.it} and Silvio De Siena \footnote{Electronic
Mail: desiena@vaxsa.dia.unisa.it}} \\

\vspace{0.2cm}

{\it Dipartimento di Fisica, Universit\`{a} di Salerno, \\
     and INFN, Sezione di Napoli, Gruppo collegato di Salerno, \\
     84081 Baronissi, Italia}

\vspace{1.1cm}

{\large Fabrizio Illuminati \footnote{Electronic Mail:
illuminati@mvxpd5.pd.infn.it}} \\

\vspace{0.2cm}

{\it Dipartimento di Fisica ``G. Galilei", Universit\`{a} di Padova, \\
     and INFN, Sezione di Padova,  \\
     35131 Padova, Italia}

\end{center}

\vspace{1cm}

{\begin{center} \large \bf Abstract \end{center}}

We show that generalized coherent states follow Schr\"{o}dinger
dynamics in time-dependent potentials. The normalized wave-packets
follow a classical evolution without spreading; in turn, the Schr\"{o}dinger
potential depends on the state through the classical trajectory. This
feedback mechanism with continuous dynamical re-adjustement
allows the packets to remain coherent indefinetely.

\vspace{0.4cm}


\newpage

Coherent states are by now of utmost importance in several areas
of theoretical physics, ranging from quantum optics to statistical
mechanics and quantum field theory \cite{klauder}. They can be
characterized as the quantum states that are closest to a classical,
localized dynamics. The problem of defining and describing such
states for general potentials traces back to Schr\"{o}dinger,
who in fact was able to give a complete, explicit solution only for the
harmonic oscillator \cite{schroedinger}.

In modern language, as originally shown by Glauber \cite{glauber},
the harmonic-oscillator
coherent states are generated by the action of the displacement
operator $D(\alpha)$ on the ground state. The same coherent states
can also be described as eigenstates of the annihilation-operator or
as the Heisenberg minimum-uncertainty states.

On the other hand, when
one tries to generalize these three methods to general non-harmonic systems,
different classes of generalized coherent states are obtained.

The displacement operator method, applied in the group-theoretical
approach, yields the most satisfying characterization
for it allows to preserve
fundamental properties of the harmonic-oscillator coherent
states such as overcompleteness and resolution of unity \cite{gilmore};
moreover, the displacement operator method can be applied without restriction
to systems with general dynamical symmetry groups \cite{gilmore},
\cite{perelomov}. In the following
we will thus focus attention only on the generalized coherent states
of the displacement operator, GCSs for brevity.

The major drawback of the
group-theoretical description of GCSs is that in this approach
it is very hard to understand the dynamical origin of their
coherent behavior and to give
an explicit description of their motion.
For instance, one would like to know
precisely in what sense they follow a classical evolution, i.e.
the form of the potential appearing in the Schr\"{o}dinger
equation obeyed by a given GCS, the form of the potential appearing in the
classical equation for the wave-packet center, and
the relation among these two and the potential associated with
the ground state being originally displaced by the action of $D(\alpha)$.

In this note we derive the explicit Schr\"{o}dinger dynamics obeyed by
GCSs. The main result of our analysis is
that GCS are always associated to time-dependent
interactions, with a back-reaction mechanism between the state
and the potential that is just what lets the packet
remain indefinetely coherent without spreading.
We sketch the proof in general
and then discuss the explicit example of the Morse potential.
Details of the general formalism and
applications to other fundamental potentials will be
presented elsewhere \cite{preparation}.

Our strategy is simply based on writing
the displacement operator $D(\alpha)$ in the
coordinate representation and then apply it to the ground state
of general non-harmonic potentials. The dynamics is then
reconstructed resorting to the Madelung hydrodynamic
representation of Schr\"{o}dinger equation.

This procedure yields non-stationary
states whose wave-packet center exactly follows
a classical equation of motion; they satisfy
Schr\"{o}dinger equation in a potential that is
time-dependent through the trajectory solution of the classical equation
for the wave-packet center. Thus the potential is subject to a back-reaction
and keeps readjusting itself through the motion
of the wave-packet center: this in turn allows the wave-packet to
keep following the classical evolution without dispersion.
In the limit of constant configurational expectation the time-dependent
potential in the Schr\"{o}dinger equation reduces to the
potential whose ground state had been originally displaced.

Finally, the potential appearing in the classical equation for the
wave-packet center is obtained as the time-independent part of the
time-dependent potential in the given GCS. For sake of simplicity,
in the following we discuss one-dimensional
systems.

Consider, in the coordinate representation, the ground state $\Psi_{0}(x)$
of a time-independent configurational potential $V(x)$.
Let the
displacement operator $D(\alpha)\,$, $\alpha(t)
= \sqrt{2\hbar}(Q(t) + iP(t))$,
act on $\Psi_{0}(x)$; one obtains the non-stationary state
$\Psi_{\alpha}(x,t)$:
\begin{equation}
\Psi_{\alpha}(x,t) \, \equiv \, D(\alpha)\Psi_{0}(x) \, = \, \exp \left( -i
\frac{PQ}{2\hbar} \right) \exp \left( i\frac{P}{\hbar}x \right)
\Psi_{0}(x - Q) \, ,
\end{equation}

\noindent where $P(t) = \langle \hat{p} \rangle_{\alpha}$,
expectation of the momentum operator in the state $\Psi_{\alpha}$,
and $Q(t) = \langle \hat{q} \rangle_{\alpha} - \langle \hat{q}
\rangle_{0}$, difference of the expectations of the
coordinate operator in the states $\Psi_{\alpha}$ and $\Psi_{0}$.
The ground-state mean $\langle \hat{q} \rangle_{0} = 0$ for symmetric
potentials; it is in general a non-zero constant for potentials with
no definite parity.

The probability density $\rho_{\alpha} = |\Psi_{\alpha}|^{2}$ is a
function of $\xi \equiv x - Q$, while the phase $S_{\alpha}$
is of the form $S_{\alpha}(x,t) = Px - PQ/2$.
In the hydrodynamic picture, to the complex
Schr\"{o}dinger equation for the state
$\Psi$ there correspond two real coupled equations for the
density $\rho$ and the phase $S$. We thus have the continuity equation
\begin{equation}
\partial_{t}\rho \; = \; -\frac{1}{m}\left[
\partial_{x}\left( \rho \partial_{x}S \right) \right] \, ,
\end{equation}

\noindent and the Hamilton-Jacobi-Madelung equation for the potential
$V$ associated to Schr\"{o}dinger dynamics:
\begin{equation}
\partial_{t}S \, + \, \frac{(\partial_{x}S)^{2}}{2m} \, - \, \frac{\hbar^{2}}
{2m}\frac{\partial_{x}^{2}\sqrt{\rho}}{\sqrt{\rho}} \; = \; - V \; .
\end{equation}

It is then easily shown by inserting
$\rho_{\alpha}$ and $S_{\alpha}$ in the above
equations that the GCS $\Psi_{\alpha}$ satisfies Schr\"{o}dinger
equation in the potential
\begin{equation}
V(x,t) \; = \; \frac{\hbar^{2}}{2m}F(\xi) \, - \, \frac{dP}{dt}x \, -
\, \frac{P^{2}}{2m} \, + \, \frac{1}{2}\left( \frac{dQ}{dt}P +
\frac{dP}{dt}Q \right) \, ,
\end{equation}

\noindent where $F(\xi) \equiv (\rho_{\alpha}(\xi))^{-1/2}d^{2}
(\sqrt{\rho_{\alpha}(\xi)})/d\xi^{2}$.

It follows from eq.(4) above that the potential depends on time through
the expectations of the observables, which in turn are determined by
the dynamical state of the system. Therefore we have proven that any
GCS follows Schr\"{o}dinger dynamics in a time-dependent potential.
The latter has the same functional form of the original time-independent
potential $V(x)$ associated to the ground state $\Psi_{0}$.
The original dynamical system is recovered
in the limit of constant classical solution $Q \to 0$.

One can also see that GCSs' wave-packet centers satisfy
\begin{equation}
\frac{dP}{dt} \; = \; - \partial_{x}V(x,t)|_{x = \langle \hat{q} \rangle } \; .
\end{equation}

\noindent This is a coherence condition: the GCSs are
driven by a classical evolution equation. The actual form of the
classical equation can be read off expanding $V$ in powers of
$x$ and equating coefficients of terms linear in $x$. This procedure
allows to identify the potential $V_{class}(x)$ entering the classical
equation as the time-independent part of the
time-dependent potential entering
Schr\"{o}dinger equation. This back-reaction of the wave-packet
motion on the Schr\"{o}dinger potential is what allows the
classical motion to hold exactly without dispersion. The price
to be paid, with respect to the harmonic-oscillator case, is that
in general $V_{class}(x)$ will not coincide with the original
potential $V(x)$ associated to $\Psi_{0}$, unless the potential
is symmetric.

We will now elucidate the general structure outlined above by
applying it to the solvable example of the Morse oscillator.
Consider the potential $V(x) = U_{0}(1-\exp[-ax])^{2}$,
where $U_{0}=\lambda^{2} {\cal{E}}_{0}$, and ${\cal{E}}_{0} =
(\hbar a)^{2}/2m$. The depth of the different
Morse wells is indexed
by the adimensional constant $\lambda > 1/2$. For computational
convenience we choose the value $\lambda =1$. The corresponding
ground state reads
\begin{equation}
\Psi_{0}(x) \; = \; \left( \frac{2\pi^{2}}{3\Delta q^{2}}
\right)^{\frac{1}{4}} \exp \left( -\gamma\frac{x}{\Delta q} -
\exp \left( -2\gamma\frac{x}{\Delta q} \right) \right) \, ,
\end{equation}

\noindent where the constant spreading $\Delta q^{2} \equiv
\langle \hat{q}^{2} \rangle_{0} - \langle \hat{q} \rangle_{0}^{2}
= 4\gamma^{2}/a^{2}$, and $\gamma=\pi/2\sqrt{6}$.
Applying $D(\alpha)$ and solving the hydrodynamic equations for
the potential yields
\begin{equation}
V(x,t) \; = \; U_{0}(1-\exp[-a\xi])^{2} \, - \, \frac{dP}{dt}x \, -
\, \frac{P^{2}}{2m} \, + \, \frac{1}{2}\left( \frac{dQ}{dt}P +
\frac{dP}{dt}Q \right) \, .
\end{equation}

\noindent From the above expression for the time-dependent
potential one readily verifies that it reduces to the time-independent
Morse well in the limit $Q \to 0$, and that condition (5)
for classical motion is satisfied.

We must now identify the classical equation obeyed by $Q(t)$, that
is the potential $V_{class}(Q)$. To this end one expands
$V(x,t)$ in powers of $x$; identifying
coefficients of the terms linear in $x$ one finally isolates the
overall linear dependence $a(t)x$ in the potential $V(x,t)$:
\begin{equation}
a(t) \; = \; -\frac{dP}{dt} \; + \; 2aU_{0}\left( \exp(aQ) \, - \,
\exp(2aQ) \right) \, .
\end{equation}

Letting the arbitrary coefficient $a(t)=0$
the classical equation for $Q(t)$ reads
\begin{equation}
\frac{dP}{dt} \; = \; 2aU_{0}\left( \exp(aQ) \, - \, \exp(2aQ) \right)
\; \equiv \; - \frac{d}{dQ}V_{class}(Q) \, ,
\end{equation}

\noindent so that the wave-packet center follows a classical motion
in the potential
\begin{equation}
V_{class}(Q) \; = \; U_{0}(1-\exp[aQ])^{2} \, .
\end{equation}

The above expression represents a repulsive Morse potential, obtained
from the original one by letting $x \to -x$. This is not surprising,
due to the non symmetric nature of the Morse oscillator. In fact,
were we to define the displacement operator with the ``wrong" sign,
$\xi \to -\xi$, then $V_{class}(x)$ would coincide with the original
Morse potential $V(x)$.
On the other hand, $V_{class}$ exactly coincides with the original
time-independent potential for
symmetric systems, such as the harmonic oscillator with centripetal
barrier, the symmetric P\"{o}schl-Teller potential and the Coulomb
potential.

Details on these systems and other results, including
the description of generalized
squeezed states and the study of time-evolution of GCSs with
non-constant dispersion, will appear elsewhere \cite{preparation}.

We mentioned in the beginning that harmonic-oscillator coherent
states are also states of minimum uncertainty. It is then possible
along this line to define a different class of generalized
coherent and squeezed states. This program was carried out by Nieto
and co-workers \cite{nieto78}, \cite{nieto93}, who considered
classically integrable systems allowing an energy-dependent
canonical transformation such that the classical Hamiltonians
are reduced, in the new ``natural" variables,
to quadratures. In this way one can quantize
a system that is formally analogous to a harmonic-oscillator
and seek the states that minimize the Heisenberg uncertainty
product written in terms of the ``natural" operators.

These minimum-uncertainty generalized coherent states present
the interesting feature of satisfying an approximate classical
motion for a finite time interval. In the light of our results
this phenomenon finds a simple explanation: for non-harmonic
systems a classical motion
can be followed indefinetely only through the feedback mechanism
driven by a time-dependent Hamiltonian.
Since Nieto's coherent states are
driven by a time-independent Hamiltonian, they eventually
become delocalized after a finite time interval.

Harmonic-oscillator coherent states can also be described
as the states that minimize the osmotic uncertainty
product in Nelson stochastic quantization \cite{demartino82}.
In the stochastic hydrodynamic picture they are states with classical
current velocity (i.e. classical phase)
and osmotic velocity linear in the process (i.e. Gaussian density)
\cite{demartino94}. If one seeks for states still with
classical phase but with densities not necessarily Gaussian,
one recovers the displacement-operator generalized coherent
states \cite{demartino94bis}. In fact, this result
inspired us to study the dynamics of GCSs in the
canonical picture.

In conclusion, we have presented the complete dynamical
characterization of generalized coherent states,
describing how to identify the classical equation
for the coherent wave-packets and
the back-reaction mechanism that is needed to
preserve coherence in the time-evolution.

In this sense generalized coherent states yield the first possible
example of state-dependent interactions in nature.
Beyond the conceptual importance in the field of quantum
coherence, this result might be useful in applications.

For instance, in the study of particle beam dynamics,
state-dependent Hamiltonians could be introduced to simulate
the experimenter's ``kicks" to keep the particle bundle coherent.
In the case of time-dependent dispersion $\Delta q(t)$, these
Hamiltonians could be of interest in the description of atomic
wave-packets with large principal quantum numbers, the
Rydberg states.

\end{document}